\documentclass{ifacconf}

\usepackage{graphicx}      
\usepackage{natbib}        

\usepackage{graphics} 
\usepackage{epsfig}
\usepackage{times} 
\usepackage{amsmath,amssymb,amsfonts}
\usepackage{mathrsfs}
\usepackage{siunitx}
\usepackage{algorithmic}
\usepackage{graphicx}
\usepackage{textcomp}
\usepackage{cuted}
\usepackage{lipsum}
\usepackage{varioref}
\usepackage{import}
\usepackage{framed,xcolor}
\usepackage{color}
\usepackage{comment}
\usepackage{multirow}
\usepackage{epstopdf}   
\usepackage{psfrag}
\usepackage{breqn}
\usepackage{float}
\usepackage{mathtools}
\usepackage{booktabs}
\usepackage{soul}
\usepackage{amsbsy}

\definecolor{seb}{rgb}{0.8,1,0.8}

\newcommand{\vect}[1]{\ensuremath{\boldsymbol{\mathrm{#1}}}}

\newtheorem{Assumption}{Assumption}

\newcommand {\G}{\textcolor{magenta}{ G }}

\usepackage{cancel}

\DeclareMathOperator*{\argmin}{arg\,min}
\begin{document}
\begin{frontmatter}

\title{Policy Gradient Reinforcement Learning for Uncertain Polytopic LPV Systems based on MHE-MPC} 


\author[First]{Hossein Nejatbakhsh Esfahani} 
\author[First]{Sébastien Gros} 

\address[First]{Department of Engineering Cybernetics, Norwegian University of Science and Technology (NTNU), Trondheim, Norway (e-mail: \{hossein.n.esfahani, sebastien.gros\}@ntnu.no).}

\begin{abstract}
In this paper, we propose a learning-based Model Predictive Control (MPC) approach for the polytopic Linear Parameter-Varying (LPV) systems with inexact scheduling parameters (as exogenous signals with inexact bounds), where the Linear Time Invariant (LTI) models (vertices) captured by combinations of the scheduling parameters becomes wrong. We first propose to adopt a Moving Horizon Estimation (MHE) scheme to simultaneously estimate the convex combination vector and unmeasured states based on the observations and model matching error. To tackle the wrong LTI models used in both the MPC and MHE schemes, we then adopt a Policy Gradient (PG) Reinforcement Learning (RL) to learn both the estimator (MHE) and controller (MPC) so that the best closed-loop performance is achieved. The effectiveness of the proposed RL-based MHE/MPC design is demonstrated using an illustrative example.
\end{abstract}

\begin{keyword}
Model Predictive Control, Moving Horizon Estimation, Reinforcement Learning, Polytopic LPV, Multi-Model Linear System 
\end{keyword}

\end{frontmatter}

\section{Introduction}
\label{sec:intro}
Model predictive control (MPC) is an advanced control approach and widely used to many control problems due to its capability to handle both input and state constraints by making explicit use of the process model to predict the future evolution of the system \citep{rawling}. However, in most applications some of the model parameters are uncertain at design time so that the prediction model may not be accurate enough to capture the true dynamics of a real system that this issue can affect the closed-loop performance \citep{CAIRANO201252}.

In many real applications of MPC, an accurate model of the (possibly nonlinear) real system may not be available so that the simplified linear models are commonly adopted as the prediction models, which are captured at a specific operation point of the real system. However, these identified models may not work for some uncertain systems where some parameters of the dynamics vary or not be known perfectly at different operation points. Therefore, the Linear Parameter-Varying (LPV) framework is usually adopted to address the issue above in which the linear models with parametric uncertainties (i.e, exogenous scheduling parameters) are represented by the LPV models \citep{LPV}.

The use of polytopic LPV prediction models has been proven to be an effective approach to develop the MPC schemes for both the linear and nonlinear systems, e.g., see \citep{LPV_MPC1,LPV_MPC2,LPV_MPC3}. In most of the MPC-based polytopic LPVs, a multi-model approach is used to represent a convex combination of the vertices as Linear Time Invariant (LTI) models in which the uncertainty (exogenous/endogenous scheduling parameter with known bounds) is associated to the convex combination vector. However, if the scheduling parameters are not known and constantly change, the robust MPC schemes, i.e, a tube MPC can be used \citep{tubeMPC}. As an alternative approach, as we consider in this paper, this combination vector is assumed to be constant or changing slowly, which can be estimated by a parameter estimation scheme \citep{10.1002,7963416}. As an interesting method to concurrently estimate this vector and some unmeasured states, a Moving Horizon Estimation (MHE) was adopted in \citep{PIPINO2020214}. However, the LTI models used as vertices of the polytopic LPV may not be exact due to the inexact selection of the scheduling parameter bounds and then, these wrong models used in both the MHE and MPC schemes can decrease the closed-loop performance. As a solution to address this problem, a learning mechanism, i.e, a Reinforcement Learning (RL) can be leveraged to learn these schemes with the aim of increasing the closed-loop performance.     

Reinforcement Learning (RL) is a powerful tool for solving Markov Decision Processes (MDP) problems \citep{sutton}, which has been recently combined with the MPC and MHE schemes in order to improve their performance where the adopted prediction/estimation models may not capture the real system dynamics perfectly, e.g., see \citep{MHEQlearning,shipACC,shipCDC}.

Motivated by this observation in the context of the MPC-based RL, in this paper, we propose to use an MHE scheme to deliver the convex combination vector needed in the MPC scheme and deal with the polytopic LPV systems with possibly inexact vertices (wrong LTI models) by adopting a Deterministic Policy Gradient (DPG) RL to adjust a certain number of the parameters in the parameterized MHE and MPC schemes. It is worth highlighting that the proposed \textit{compatible deterministic policy gradient reinforcement learning based on a combined MHE-MPC scheme} is a central contribution in this paper. Although this learning-based (state-parameter) estimator/controller scheme is developed and employed in the context of the LPV systems in this paper, the proposed algorithm can be readily used to control many other applications, i.e, partially observable systems in which a combined estimator/observer-MPC is needed.   

In what follows, the polytopic LPV model is introduced in Section \ref{sec:LPV}, and the parameterized MHE and MPC schemes are detailed in Section \ref{sec:MHE/MPC}. The MHE/MPC-based policy gradient reinforcement learning is then developed in Section \ref{sec:RL}. Simulations on a mass-spring-damper case study are reported in Section \ref{sec:exm}, and finally, conclusions and future work are given in Section \ref{sec:con}.
\section{Uncertain Polytopic LPV}\label{sec:LPV}
In this paper, we focus on the classic (pure) LPV systems, in which the uncertain dynamical model of the system can be described as follows:
\begin{subequations}
\label{eq::LPV-SS}
\begin{align}
    &\vect x_{k+1}=A\left(\vect\rho_k\right)\vect x_k+B\left(\vect\rho_k\right)\vect u_k\\
    &\vect y_k=C(\vect\rho_k)\vect x_k
\end{align}
\end{subequations}
where the scheduling parameters $\vect\rho\in\mathcal{P}\in\mathbb{R}^p$ are considered as some exogenous signals applied to the real system. The vectors $\vect x\in \mathbb{R}^n$, $\vect u\in \mathbb{R}^m$ are states and control inputs, respectively. The system measurements is labeled by $\vect y_k\in\mathbb{R}^e$. Hence, the system \eqref{eq::LPV-SS} can be represented by a polytopic approximation $\left[A\left(\vect\rho_k\right),B\left(\vect\rho_k\right),C\left(\vect\rho_k\right)\right]\in\Omega$, where a polytope $\Omega$ is defined as a convex hull upon the Linear Time Invariant (LTI) models as follows: 
\begin{align}
    \Omega=Co\left\{[A_1,B_1,C_1],[A_2,B_2,C_2],\cdots,[A_\ell,B_\ell,C_\ell]\right\}
\end{align}
where $\ell=2^p$ is the number of vertices and $Co\{\cdot\}$ reads a convex hull. Indeed, the LTI matrices are delivered by combinations of the scheduling parameters at their bounds $\underline{\vect\rho},\bar{\vect\rho}$. We then evaluate the variable matrices in \eqref{eq::LPV-SS} as a convex combination of the LTI models (a.k.a vertices) as follows:
\begin{subequations}
   \begin{align}
         &A\left(\vect\rho_k\right)=\sum_{i=1}^{\ell}\beta_{i}A_i,\quad B\left(\vect\rho_k\right)=\sum_{i=1}^{\ell}\beta_{i}B_i,\\
         &C\left(\vect\rho_k\right)=\sum_{i=1}^{\ell}\beta_{i}C_i
   \end{align}
\end{subequations}
where $A_i,B_i,C_i$ are known matrices of appropriate size and $\beta_{i},i=1\cdots\ell$ is the unknown weight of each vertex (a.k.a convex combination vector) such that:
\begin{align}\label{eq::beta}
    \chi=\left\{\vect\beta\in\mathbb{R}^\ell:\sum_{i=1}^{\ell}\beta_{i}=1,\quad 0\leq\beta_{i}\leq 1\right\}
\end{align}
\begin{Assumption}
By \citep{7525467}, the convex combination vector $\vect\beta_k$ is assumed to be constant or changing slowly with respect to the dynamics and there exists an estimator to compute the estimate of this vector such that $\vect{\hat\beta}_k\in\chi$ for all $k\in\mathbb{Z}_{0+}$.
\end{Assumption}
Note that the stability and constraint satisfaction of the MPC schemes for the polytopic LPV systems under Assumption 1 was detailed and established in \citep{10.1002} and by Lemma 2 and Theorem 1 in \citep{7525467}. We then focus on this kind of the LPV system where its vertices may not be known and exact. As discussed, the vertices in a polytopic LPV may have a wrong model because of the inexact bounds of the scheduling parameters, some parametric uncertainties and the linearization approximation effects. Then, the wrong LTI models in the polytope $\Omega$ can be introduced by the inexact vertices as follows:
\begin{align}
    \hat\Omega=Co\left\{[\hat A_1,\hat B_1,\hat C_1],[\hat A_2,\hat B_2,\hat C_2],\cdots,[\hat A_\ell,\hat B_\ell,\hat C_\ell]\right\}
\end{align}
In the context of MPC schemes for the polytopic LPVs, under Assumption 1, one needs to evaluate an estimation of $\beta_{i},i=1\cdots\ell$ \citep{10.1002,PIPINO2020214}. In this paper we propose to adopt an MHE scheme to deliver these unknown weights  while some unmeasured states are estimated. However, the inexact vertices in $\hat\Omega$ can affect parameter/state estimations, the policy captured from the MPC and the closed-loop performance, subsequently. In addition to this inexact polytope issue, the partially measurable states can be challenging since the MHE scheme as an estimator of $\beta_{i,k}$ needs to capture all states as measurements and minimize a distance as a model matching error between these measurements and those states delivered by the uncertain LPV model.

To cope with these problems, we propose to parameterize both the MHE and MPC schemes and adopt a reinforcement learning to adjust them in order to achieve the best closed-loop performance even when using some wrong LTI models in a polytopic LPV system. More precisely, it is possible to generate the optimal policy based on a wrong model by learning the MHE and MPC schemes. 
\section{Adjustable MHE-MPC in a LPV Framework}\label{sec:MHE/MPC}
Under Assumption 1, an obvious choice for the combination vectors used in the MPC scheme would be $\vect\beta_{j|k}=\vect\beta_k$, for all $j\in\mathbb{Z}_{k,\ldots,k+N}$. As discussed in \citep{7525467}, this choice could not be proper because if the entire $\beta$-parameter prediction vector suddenly changes, the MPC value function may not be decreasing. Hence, under Assumption 1, one can select a $N$-step delay in this parameter prediction as follows:
\begin{align}
    \vect\beta_{j|k}=\vect{\hat\beta}_{k-N+j},\quad \forall j\in\mathbb{Z}_{0,\ldots,N}
\end{align}
We then adopt an MHE-based estimator to deliver these weights needed in an MPC scheme. Let us formulate a parameterized MHE scheme as $\beta$-parameter and state estimator, where the vertices of the polytopic LPV are inexact and the real system states are not fully measurable. 
\begin{subequations}\label{eq:mhe}
		\begin{align}
			&\left \{ \vect{\hat\beta}_{k-N\ldots k},\hat{\vect x}_{k-N\ldots k} \right \}=\argmin_{\vect\beta,\vect x}\gamma^{N}
			\begin{Vmatrix}
                    \vect x_{k-N}-\tilde{\vect x}_{k-N}
            \end{Vmatrix}_{A_\theta}^2 \nonumber\\
			&\qquad+\sum_{j=k-N+1}^{k}\gamma^{k-j}\left(\left \|\vect\mu_j \right \|_{R_\theta}^2 +\left \|\Delta\vect\beta_{j|k}\right \|^2\right)\nonumber\\
			&\quad\mathrm{s.t.} \quad \vect x_{j+1}=\hat A\left(\vect\rho_j\right)\vect x_j+\hat B\left(\vect\rho_j\right)\bar{\vect u}_j,\\
			&\quad\hat A\left(\vect\rho_j\right)=\sum_{i=1}^{\ell}\beta_{j,i}\hat A_i,\quad \hat B\left(\vect\rho_j\right)=\sum_{i=1}^{\ell}\beta_{j,i}\hat B_i,\\
			&\quad\bar{\vect y}_j=\left(\sum_{i=1}^{\ell}\beta_{j,i}\hat C_i\right)\vect x_j+\vect\mu_j,\\
			&\quad\sum_{i=1}^{\ell}\beta_{j,i}=1,\quad 0\leq\beta_{j,i}\leq 1\\
			&\quad\Delta\vect\beta_{j|k}=\vect\beta_{j|k}-\vect\beta_{j|k-1},\quad\vect\beta_j=Col\left\{\beta_{j,1},\ldots,\beta_{j,\ell}\right\}
		\end{align}
\end{subequations}
where $\gamma\in\mathbb{Z}_{(0,1)}$ is a discount factor and $\bar{\vect y}_j,\bar{\vect u}_j$ are the measurements available at the physical time $k$. We label $R_\theta$ as an adjustable weight matrix penalizing the deviation between some estimated states captured from the LPV model and their corresponding measurements. The first term in the above least squares problem is an arrival cost weighted with matrix $A_\theta$, which aims at approximating the information prior to $k-N_{\text{MHE}}$, where $\tilde{\vect x}$ is the available estimation $\hat{\vect x}_{k-N}$ at time $k-1$. We then detail a policy gradient RL in order to adjust the combination vectors $\beta_{i}$, where the LTI models (vertices of the polytope) are wrong.

This paper proposes a learning-based estimation and control for a polytopic LPV system with possibly inexact LTI models at the vertices, where the control policy is delivered by a parameterized MPC scheme as follows:
\begin{subequations}\label{eq:mpc}
		\begin{align}
			&\min_{\vect x,\vect u,\vect \sigma }\quad \gamma^{k+N}\left(T_{\theta}(\vect x_{k+N},\vect\beta_{k+N})+\vect w_{f}^\top\vect\sigma_{k+N}\right) \nonumber\\
			&+\sum_{j=k}^{k+N-1}\mathcal{G}_\theta(\vect x_j,\vect u_j)+\sum_{j=k}^{k+N-1}\gamma^{j}\left(l_{\theta}(\vect x_{j},\vect u_{j})+\vect w^\top\vect \sigma_{j}\right)\label{eq:v1}\\
			\mathrm{s.t.}
			&\quad \vect x_{j+1}=\hat A\left(\vect\rho_j\right)\vect x_j+\hat B\left(\vect\rho_j\right)\vect u_j,\\
			&\quad\vect x_k=\hat{\vect x}_k,\quad \vect\beta_{j|k}=\vect{\hat\beta}_{j-N}\\
			&\quad\hat A\left(\vect\rho_j\right)=\sum_{i=1}^{\ell}\beta_{j,i}\hat A_i,\\
			&\quad\hat B\left(\vect\rho_j\right)=\sum_{i=1}^{\ell}\beta_{j,i}\hat B_i,\\
			&\quad \vect g(\vect u_{j})\leq 0,\\
			&\quad \vect h_{\theta}(\vect x_{j},\vect u_j)\leq \vect \sigma_{j},\quad \vect h_{\theta}^{f}(\vect x_{k+N})\leq \vect \sigma_{k+N}  \label{eq:violation}\\
			&\quad \vect\sigma_{k,\ldots,k+N} \geq 0 
		\end{align}
\end{subequations}
where $T_\theta$ and $l_\theta$ are the parameterized terminal and stage costs, respectively. The function $\mathcal{G}_\theta$ is labeled the cost modification term, which can be, i.e, in a form of gradient $\vect f^{\top}[\vect x_j,\vect u_j]^{\top}$. However, the polytopic LPV model is uncertain with wrong vertices that could cause some violations in its constraints and bring the MPC in an infeasible region. To avoid such an infeasibility, we introduce some slack variables $\vect\sigma$ penalized by enough large weights $\vect w,\vect w_f$. We also propose to have some adjustable parameters upon inequality constraints $h_\theta,h^f_\theta$. Then, we let RL to tune the MPC parameters (a.k.a policy parameters). Note that, the proposed parameterization of the MHE and MPC is a general form of a rich parameterization and one may take into account some of these parameters to be adjusted by RL.

Note that the parameterized terminal cost in the MPC scheme \eqref{eq:mpc} is defined as a function of the terminal states and the convex combination vector as follows:
\begin{align}
    T_\theta=\vect x_{k+N}^\top\left(\sum_{i=1}^{\ell}\beta_{k+N,i} P_i\right)\vect x_{k+N}
\end{align}
where the matrices $P_i>0,i\in\mathbb{Z}_{1,\ldots,\ell}$ are adjusted by RL. However, the initial values for these matrices can be computed offline via a Linear Matrix Inequality (LMI) based on a parameter-dependent Lyapunov function and a parameter-dependent linear control law detailed in \citep{10.1002}. 
\section{MHE/MPC-based Policy Gradient RL}\label{sec:RL}
Let us define the closed-loop performance of a parameterized policy $\vect\pi_\theta$ for a given stage cost $L\left(\vect x,\vect a\right)$ as the following total expected cost:
\begin{align}
    J\left(\vect\pi_\theta\right)=\mathbb{E}_{\varrho_s}\Bigg[\sum_{k=0}^{\infty}\gamma^kL\left(\vect x_k,\vect a_k\right)\Bigg|\vect a_k=\vect \pi_\theta(\vect x_k)\Bigg]
\end{align}
where the expectation $\mathbb{E}_{\varrho_s}[\cdot]$ is taken over the distribution of the Markov chain $\varrho_s$ in the closed-loop under policy $\vect\pi_\theta$. We then seek the optimal policy parameters as follows:
\begin{align}
    \vect\theta_\star=\mathrm{arg}\min_{\vect\theta}J\left(\vect\pi_\theta\right)
\end{align}

As a learning-based control approach in this paper, we propose to use a parameterized MPC scheme as a policy approximation in order to deliver $\vect\pi_\theta$. As an advantage to use a MPC-based policy approximation instead of DNNs, it can offer a learning-based controller so that all the state-input constraints are satisfied and the closed-loop performance is improved by adjusting the policy parameters.

For a given estimated state $\hat{\vect x}_{k}$ and parameter $\vect{\hat\beta}_k$ delivered by the MHE scheme, the policy captured by a parameterized MPC scheme is
\begin{align}
    \vect\pi_\theta\left(\hat{\vect x}_{k},\vect{\hat\beta}_k\right)=\vect u_0^\star\left(\hat{\vect x}_{k},\vect{\hat\beta}_k,\vect\theta\right)
\end{align}
where $\vect u_0^\star$ is the first element of the control input sequence $\vect u^\star$ delivered by \eqref{eq:mpc}.

In this section, we propose a policy gradient RL framework based on MPC and MHE to deal with the polytopic LPV systems with wrong LTI models.  
\label{sec:DPG}
\subsection{Compatible Deterministic Actor-Critic }
In the context of DPG-based RL algorithms, the policy parameters $\vect\theta$ can be directly optimized by the gradient descent steps such that the best expected closed-loop cost (a.k.a policy performance index $J$) can be captured by applying the policy $\vect\pi _{\vect\theta}$.
\begin{align}
\label{eq:theta}
    \vect\theta \leftarrow \vect\theta-\alpha  \nabla _{\vect\theta}J(\vect\pi _{\vect\theta})
\end{align}
for some $\alpha>0$ small enough as the step size. In the context of hybrid controller/observer scheme MHE-MPC, the input signal can be interpreted as a sequence of measurements $\vect y=\left[y^1_{0,\cdots,k},\cdots,y^{n_y}_{0,\cdots,k}\right]^\top$ at the physical time $k$, where $n_y$ is the number of system states selected as measurable outputs. Then, the intermediate variables $\hat{\vect x}_{k},\vect{\hat\beta}_k$ are delivered by the MHE scheme based on the history of the measurements and fed to the MPC scheme to deliver the control policy. Assuming that this history constitutes a Markov state, one can define a policy performance index by the following expected value:
\begin{align}\label{eq:j}
    J(\vect\pi _{\vect\theta}):= \mathbb{E}_{{\vect { \pi}_{\vect\theta}}}\left[Q_{{\vect\pi _{\vect\theta} }}\left(\hat{\vect x}_{k},\vect\pi_\theta\left(\hat{\vect x}_{k},\vect{\hat\beta}_k\right)\right)\right]
\end{align}
where the expectation $\mathbb{E}_{{\vect { \pi}_{\vect\theta}}}$ is taken over the distribution of the Markov chain resulting from the real system in closed-loop with ${\vect{ \pi}}_{\vect{\theta}}$. The action-value function $Q_{\vect{\pi}_{\vect{\theta}}}$ is then defined as follows:
\begin{align}
{Q_{\vect\pi _{\vect\theta }}}\left( {\hat{\vect x}_{k},\vect a_k} \right) &= L\left( {\vect y_k,\vect a_k} \right) + \gamma \mathbb{E}_{{\vect { \pi}_{\vect\theta}}}\left[ {{V_{\vect\pi _{\vect\theta }}}\left({\hat{\vect x}_{k+1} }\right)|\hat{\vect x}_{k},\vect a_k} \right] 
\end{align}
where, $L\left( {\vect y_k,\vect a_k} \right)$ reads as baseline cost (RL stage cost), which is a function of measurable states and actions at the current time $k$. Based on the proposed DPG theorem by \citep{silver2014deterministic} and the fact that both the $\vect\pi _{\vect\theta }$ and $Q_{\vect\pi _{\vect\theta }}$ are functions of $\hat{\vect x}_{k}$, the policy gradient equation is described as follows:
\begin{align}\label{eq:dJ1}
    \nabla _{\vect\theta}J(\vect\pi _{\vect\theta}) = \mathbb E\left[{\nabla _{\vect\theta} }{\vect\pi _{\vect\theta} }(\hat{\vect x}_{k}){\nabla _{\vect u}}{Q_{{\vect\pi _{\vect\theta} }}}(\hat{\vect x}_{k},\vect a_k)|_{\vect a_k=\vect \pi _{\vect\theta}}\right]
\end{align}
To represent the effect of the parameterized MHE upon the policy gradient, the sensitivity of the policy w.r.t $\vect\theta$ can be updated such that the new policy gradient is described by the following expectation:
\begin{align}\label{eq:dJ2}
    \nabla _{\vect\theta}J(\vect\pi _{\vect\theta})=\mathbb E\Big[\Xi{\nabla _{\vect u}}{Q_{{\vect\pi _{\vect\theta} }}}(\hat{\vect x}_{k},\vect a_k)|_{\vect a_k=\vect \pi _{\vect\theta}}\Big]
\end{align}
where the Jacobian matrix $\Xi$ is obtained by the following chain rule:
\begin{align}
    \Xi=\left(\nabla _{\vect\theta} \vect\pi _{\vect\theta}+\nabla_{\vect\theta}\hat{\vect x}_{k}\nabla_{\hat{\vect x}_{k}}\vect\pi _{\vect\theta}+\nabla_{\vect\theta}\vect{\hat\beta}_k\nabla_{\vect{\hat\beta}_k}\vect\pi _{\vect\theta}\right)
\end{align}
In this paper, we adopt a \textit{compatible deterministic actor-critic} algorithm \citep{silver2014deterministic} in which the action-value function $Q_{\vect\pi _{\vect\theta}} (\hat{\vect x}_{k},\vect a_k)$ can be replaced by a class of compatible function approximator $Q^{\vect w}(\hat{\vect x}_{k},\vect a_k)$ such that the policy gradient is preserved. Therefore, the compatible function for a deterministic policy $\vect\pi _{\vect\theta}$ delivered by the parameterized MHE-MPC scheme can be expressed as follows:
\begin{align}
\label{eq:Q_w}
&Q^{\vect w}={{\left( {\vect a_k - {\vect\pi _{\vect\theta} }} \right)}^{\top}}\Xi^{\top}{\vect w} + { V^{\vect\nu}}\left( \hat{\vect x}_{k}\right)
\end{align}
The first term in the above compatible function as critic part is an estimation for the advantage function and the second term estimates a value function for the history of the measurements delivered as a summarized variable $\hat{\vect x}_{k}$ by the MHE scheme. Both functions can be computed by the linear function approximators as follows:
\begin{subequations} 
	\begin{align}
       &{V^{\vect\nu}} \left(\hat{\vect x}_{k}\right ) =\vect\Upsilon\left(\hat{\vect x}_{k} \right)^\top {\vect \nu},\\
       &{A^{\vect w}} \left(\hat{\vect x}_{k},\vect a_k\right ) =\vect \Psi\left(\hat{\vect x}_{k},\vect a_k \right)^\top {\vect w}
	\end{align}
\end{subequations}
where $\vect\Upsilon\left(\hat{\vect x}_{k}\right)$ is the summarized measurement feature vector in order to constitute all monomials of the history of the measurements with degrees less than or equal to $2$. The vector $\vect \Psi\left(\hat{\vect x}_{k},\vect a_k \right):=\Xi\left( {\vect a_k - \vect\pi_\theta(\hat{\vect x}_{k},\vect{\hat\beta}_k)} \right)$ includes the state-action features. Considering \eqref{eq:Q_w}, the policy gradient \eqref{eq:dJ2} is then rewritten as follows: 
\begin{align}\label{eq:dJ3}
    &\nabla _{\vect\theta}J(\vect\pi _{\vect\theta})=\mathbb E\left[ \Xi\Xi^\top {\vect w}\right]
\end{align}
In this paper, the parameterized policy in the context of policy gradient RL is proposed to be captured by the MPC scheme \eqref{eq:mpc}. To evaluate the policy gradient \eqref{eq:dJ3}, one need to calculate some sensitivities upon the MPC and MHE schemes in order to compute the Jacobian matrix $\Xi$. Hence, the Jacobian matrices ${\nabla _{\vect\theta} }{\vect\pi _{\vect\theta} }$, $\nabla_{\hat{\vect x}_{k}}\vect\pi _{\vect\theta}$ and $\nabla_{\vect{\hat\beta}_k}\vect\pi _{\vect\theta}$ can be computed by the sensitivity analysis for the parameterized MPC scheme while the gradients ${\nabla _{\vect\theta} }\hat{\vect x}_{k}$ and $\nabla_{\vect\theta}\vect{\hat\beta}_k$ are obtained as a sensitivity term for the parameterized MHE scheme. 
\subsection{Sensitivity Analysis and LSTD-based DPG}
\subsubsection{Sensitivity Computation}
We describe next how to compute the sensitivities (gradients) needed in the proposed policy gradient RL framework based on MHE-MPC. To that end, let us define the Lagrange functions $\hat{\mathcal{L}}_\theta,\mathcal{L}_\theta$ associated to the MHE and MPC schemes \eqref{eq:mhe}, \eqref{eq:mpc} as follows:
	\begin{gather}
		\hat{\mathcal{L}}_{\theta}\left(\hat z\right)=\hat{\Lambda}_{\theta}+\hat{\vect \lambda}^\top \hat{G}_{\theta}\\
		\mathcal{L}_{\theta}\left(z\right)=\Lambda_{\theta}+\vect \lambda^\top G_{\theta}+\vect \mu^\top H_{\theta}
	\end{gather}
	
	where $\Lambda_\theta$ and $\hat{\Lambda}_\theta$ are the total parameterized costs of the MPC and MHE schemes, respectively. The inequality constraints of \eqref{eq:mpc}  are collected by $H_{\theta}$ while $G_\theta$ and $\hat{G}_\theta$ gather, respectively, the equality constraints in the MPC and MHE schemes. We then label $\vect \lambda, \hat{\vect \lambda}$ the Lagrange multipliers associated to the equality constraints $G_\theta,\hat{G}_\theta$ of the MPC and MHE, respectively. Variables $\vect\mu$ are the Lagrange multipliers associated to the inequality constraints of the MPC scheme. Let us label $\vect\Gamma=\left\{\vect x,\vect u, \vect\sigma\right\}$ and $\hat{\vect\Gamma}=\left\{\hat{\vect x},\hat{\vect\beta}\right\}$ the primal variables for the MPC and MHE, respectively. The associated primal-dual variables then read as $\vect z=\left\{\vect\Gamma,\vect \lambda,\vect\mu\right\}$ and $\hat{\vect z}=\left\{\hat{\vect\Gamma},\hat{\vect \lambda}\right\}$.
	
	The sensitivity of the policy delivered by the MPC scheme \eqref{eq:mpc} w.r.t policy parameters and the sensitivity of the estimated state associated to  the  MHE  scheme \eqref{eq:mhe} can be obtained via using the Implicit Function Theorem (IFT) on the Karush Kuhn Tucker (KKT) conditions underlying the parametric NLP. Assuming that Linear Independence Constraint Qualification (LICQ) and Second Order Sufficient Condition (SOSC) hold \citep{nlp} at $\vect z^\star$ and $\hat{\vect z}^\star$, then, the following holds:  		
	\begin{subequations}\label{eq:Sens}
	\begin{align}
		&\frac{\partial \vect z^\star}{\partial \theta}=-\frac{\partial \kappa_\theta}{\partial \vect z}^{-1}\frac{\partial \kappa_\theta}{\partial \theta},\\
		&\frac{\partial \hat{\vect z}^\star}{\partial \theta}=-\frac{\partial \hat \kappa_\theta}{\partial \hat{\vect z}}^{-1}\frac{\partial \hat \kappa_\theta}{\partial \theta}
	\end{align}
	\end{subequations}
	where 
	\begin{align}
		\kappa_{\theta}=\begin{bmatrix}
			\nabla_{\vect\Gamma}\mathcal{L}_{\theta}\\G_{\theta} \\{\mathrm{diag}\left(\vect\mu\right) \vect H_{\vect\theta} }
		\end{bmatrix}, \quad \hat \kappa_{\theta}=\begin{bmatrix}
			\nabla_{\hat {\vect\Gamma}}\hat{\mathcal{L}}_{\theta}\\\hat{G}_{\theta} 
		\end{bmatrix}
	\end{align}
are the KKT conditions associated to the MPC and MHE schemes, respectively.
As $\vect\pi_\theta$ and $(\hat{\vect x},\hat{\vect\beta})$ are, respectively, part of $\vect z^\star$ and $\hat{\vect z}^\star$. Then, the sensitivity of the MPC policy ${\nabla _{\vect\theta} }{\vect\pi _{\vect\theta} }$ and the sensitivity of the MHE solution  $(\nabla_{\vect\theta}\hat{\vect x},\nabla_{\vect\theta}\hat{\vect\beta})$ required in \eqref{eq:dJ3} can be extracted from gradients $\frac{\partial \vect z^\star}{\partial \theta}$ and $\frac{\partial \hat{\vect z}^\star}{\partial \theta}$, respectively.
\subsubsection{LSTD-based Policy Gradient}
In the context of compatible DPG, one can evaluate  the optimal parameters $\vect w$ and $\vect\nu$ of the action-value function approximation \eqref{eq:Q_w} as solutions of the following Least Squares (LS) problem:
\begin{align}
\label{eq:error}
    \min_{{\vect w}, \vect\nu} \mathbb{E} \left[\big( Q_{\vect\pi_{\vect\theta}}(\hat{\vect x}_{k},\vect a_k)-Q^{\vect w}(\hat{\vect x}_{k} ,\vect a_k)\big )^2\right],
\end{align}
In the context of RL, the LSTD-based algorithms offer an efficient use of data and tend to converge faster than other methods, e.g., see \citep{esfahani2021approximate}. The LSTD update rules for a policy gradient RL are then obtained as follows:
\begin{subequations}
\begin{align}
    &\vect\nu=A_{\vect\nu}^{-1}b_{\vect\nu},\\
    &{\vect w}=A_{\vect w}^{-1}b_{\vect w},\\
    &\vect\theta\leftarrow {\vect\theta} -\alpha b_{\vect\theta}
\end{align}
\end{subequations}
where the matrices $\Omega_{(\cdot)}$ and the vectors $b_{(\cdot)}$ are calculated by taking expectation ($\mathbb{E}_m$) over $m$ episodes as follows:
\begin{subequations}
\small
\label{eq:lstdq_update}
\begin{align}
    &A_{\vect\nu} = \mathbb E_{m} {\left[ {\sum_{k=1}^{T_f}\left[ {\vect\Upsilon\left(\hat{\vect x}_{k} \right){{\left( {\vect\Upsilon\left(\hat{\vect x}_{k} \right) - \gamma \vect\Upsilon\left(\hat{\vect x}_{k+1} \right)} \right)}^\top}} \right]} \right]}\\
    &A_{\vect w}=\mathbb E_{m} \left[ {{\sum_{k=1}^{T_f}}\Big[ \vect \Psi\left(\hat{\vect x}_{k},\vect a_k \right)\vect \Psi\left(\hat{\vect x}_{k},\vect a_k \right)^{\top} \Big]} \right],\\
    &b_{\vect\nu}=\mathbb E_{m}\left[\sum_{k=1}^{T_f}{\vect\Upsilon\left(\hat{\vect x}_{k} \right)L(\vect y_k,\vect a_k)} \right],\\
    &b_{\vect w}=\\\nonumber
    &\mathbb E_{m}\left[\sum_{k=1}^{T_f}\Big[ {\left( {L(\vect y_k,\vect a_k) + \gamma V^{\vect\nu}\left( {{\hat{\vect x}_{k+1}}} \right) - V^{\vect\nu}\left( \hat{\vect x}_{k} \right)} \right)\vect\Psi(\hat{\vect x}_{k},\vect a_k)} \Big]\right],\\
    &b_{\vect\theta}=E_{m} \Bigg[\sum_{k=1}^{T_f}\Xi\Xi^\top \vect w \Bigg]
\end{align}
\end{subequations}
where $T_f$ is the final time instant at the end of each episode.

\section{Illustrative Example}\label{sec:exm}
\begin{figure}[htbp!]
		\centering
		\includegraphics[width=.8\linewidth]{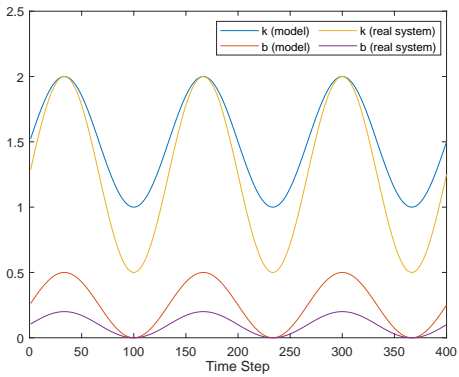}
		\caption{Mass-Spring-Damper with variable spring constant and damping factor as scheduling parameters} 
		\label{sch}
\end{figure}

\begin{figure}[htbp!]
		\centering
		\includegraphics[width=1\linewidth]{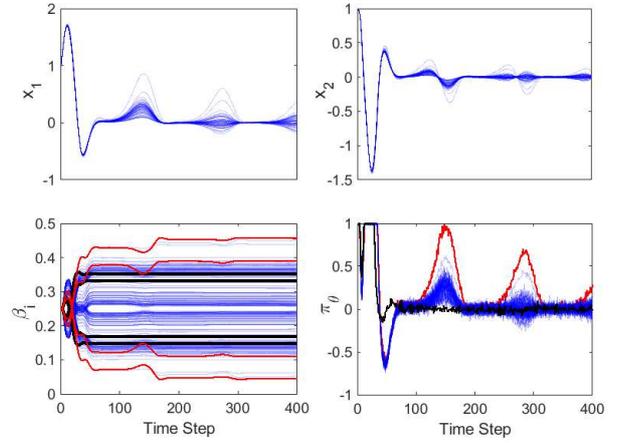}
		\caption{Evolution during reinforcement learning: states, policy and $\beta$-parameters} 
		\label{evolution}
\end{figure}

\begin{figure}[htbp!]
		\centering
		\includegraphics[width=1\linewidth]{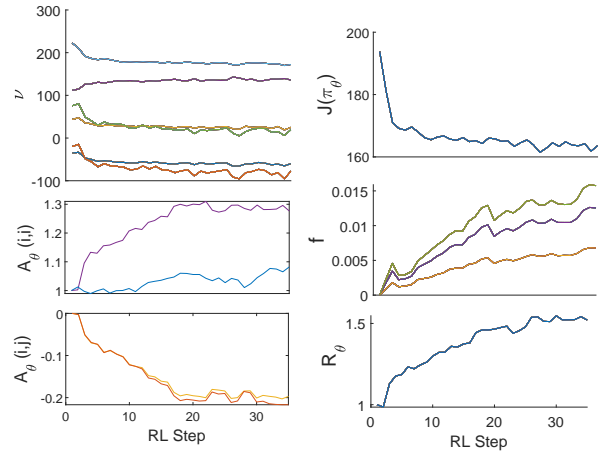}
		\caption{Closed-loop performance and parameters of the policy gradient reinforcement learning} 
		\label{RL}
\end{figure}

\begin{figure}[htbp!]
		\centering
		\includegraphics[width=1\linewidth]{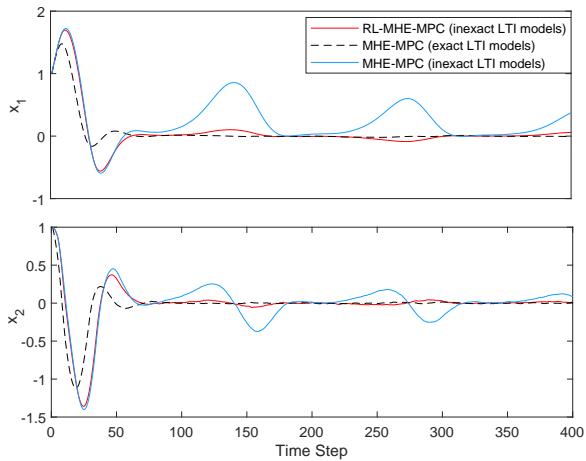}
		\caption{Comparative analysis} 
		\label{comp}
\end{figure}

To demonstrate the effectiveness of the proposed learning-based control approach, we choose a stabilization problem for a mass-spring-damper system where the spring and damper coefficients are considered as the exogenous scheduling parameters.
\begin{align}
    \frac{d}{dt}\begin{bmatrix}x_1\\x_2\end{bmatrix}=\begin{bmatrix}0&1\\-k/m&-d/m \end{bmatrix}\begin{bmatrix}x_1\\x_2\end{bmatrix}+\begin{bmatrix}0\\1/m\end{bmatrix}u
\end{align}
Then, the LPV model can be represented as a linear system with polytopic uncertainty including four vertices (LTI models), which are constructed based on the scheduling bounds $1\leq k\leq 2,0\leq b\leq 0.5$ shown in Figure \ref{sch}. 

\begin{align}
     &A_1=\begin{bmatrix}0&1\\-1&0 \end{bmatrix},\quad A_2=\begin{bmatrix}0&1\\-1&-0.5 \end{bmatrix}\\\nonumber
     &A_3=\begin{bmatrix}0&1\\-2&0 \end{bmatrix},\quad A_4=\begin{bmatrix}0&1\\-2&-0.5 \end{bmatrix}
\end{align}
and $B_{1,\ldots,4}=\left[0,1\right]^\top$. On the other hand, we select the scheduling bounds used in the real system different from those adopted in four LTI models $0.5\leq k\leq 2,0\leq b\leq 0.2$. The position of the mass ($x_1$) is selected as measurement while the velocity ($x_2=\dot x_1$) is estimated. We consider a constraint on the control input $-1\leq u\leq 1$.
As it can be observed in Figure \ref{evolution}, the stabilization scenario is affected since the scheduling parameters violate their correct values used in four LTI models. We then adopt a policy gradient RL to adjust the MHE-MPC parameters in order to cope with this problem and improve the closed-loop performance $J(\pi_\theta)$. The evolution of some RL parameters and the closed-loop performance are shown in Figure \ref{RL}. The evolution during learning shows that the optimal policy (Figure \ref{evolution}: $\pi_\theta$ in black color) can be delivered after $35$ RL step and the stabilization goal is perfectly achieved. Note that the system's behavior in terms of $\beta$-parameters and control policy without learning is highlighted in red color shown in Figure \ref{evolution}. It is obvious that the estimated combination vector $\vect\beta$ is improved by adjusting the MHE scheme under wrong LTI models. As a comparative study shown in Figure \ref{comp}, the performance of the stabilization has been dramatically improved after learning the MHE-MPC scheme where the LTI models cannot capture the correct polytope based on the scheduling bounds used in the real system.  

\section{Conclusion}\label{sec:con}
In this paper, an MHE-MPC scheme was proposed to deal with the linear systems with polytopic uncertainty in the context of the LPV systems with exogenous scheduling parameters. However, the performance of the adopted multi-model approach can be affected where the bounds of the scheduling parameters are different from their real values applied to the real system. To address a data-driven MHE-MPC approach for the polytopic LPV systems with inexact LTI models, we proposed to adopt a policy gradient reinforcement learning to capture the optimal policy and achieve the best closed-loop performance by adjusting the MHE and MPC schemes.  

\begin{ack}
This work was supported by the Research Council of Norway (RCN), project ¨Safe Reinforcement Learning using MPC (SARLEM)¨. 
\end{ack}

\bibliography{ifacconf}            
\end{document}